\documentclass[12pt,preprint]{aastex}

\newcommand{\kms}{\rm km\ s^{-1}}
\newcommand{\HI}{H\,{\small I}}
\newcommand{\HII}{H\,{\small II}}

\shorttitle{Compact \HI\ clouds}
\shortauthors{Stil et~al.}

\begin{document}

\title{Compact \HI\ clouds at high forbidden velocities in the inner Galaxy}

\author{J. M. Stil\altaffilmark{1}}
\author{Felix. J. Lockman\altaffilmark{2}}
\author{A. R. Taylor\altaffilmark{1}}
\author{J. M. Dickey\altaffilmark{3,4} and D. W. Kavars\altaffilmark{3}}
\author{P. G. Martin, T. A. Rothwell, and A. Boothroyd\altaffilmark{5}}
\author{N. M. McClure-Griffiths\altaffilmark{6}}
\altaffiltext{1}{Department of Physics and Astronomy, University of Calgary, 2500 University Drive N.W., Calgary AB T2N 1N4, Canada}
\altaffiltext{2}{National Radio Astronomical Observatory, P.O. Box 2, Green Bank, West Virginia 24944, USA}
\altaffiltext{3}{Department of Astronomy, University of Minnesota, 116 Church Street, SE, Minneapolis, MN 55455, USA}
\altaffiltext{4}{School of Mathematics and Physics - Private Bag 37, University of Tasmania, Hobart, TAS 7001, Australia}
\altaffiltext{5}{Department of Astronomy and Astrophysics, University of Toronto, 60 St. George Street Room 1403, Toronto ON, M5S 3H8, Canada}
\altaffiltext{6}{Austalia Telescope National Facility, CSIRO, P.O. Box 76, Epping NSW 1710, Australia}
\begin{abstract}

The VLA Galactic Plane Survey (VGPS) of the first Galactic quadrant
was searched for \HI\ emission with velocities well above the maximum
velocity allowed by Galactic rotation. A sample of 17 small
fast-moving clouds was identified. The distribution of the ensemble of
clouds in longitude and velocity indicates that the clouds are part of
the Galactic disk, despite their large forbidden velocity. The median
angular diameter of the clouds detected in the VGPS is
$3\farcm4$. These clouds would not be noticed in previous low
resolution surveys because of strong beam dilution.  Assuming each
cloud is located at the tangent point, a median cloud has a diameter
of 10 pc, \HI\ mass of $60\ M_\odot$, and a velocity more than $25\
\kms$ beyond the local terminal velocity derived from $\rm^{12}CO$
observations. Three clouds in the sample have a velocity between $50$
and $60\ \kms$ in excess of the local terminal velocity.  The
longitude distribution of the sample peaks near $l = 30^\circ$, while
the latitude distribution of the clouds is nearly flat. The observed
longitude and latitude distributions are compared with simulated
distributions taking into account the selection criteria of the cloud
search. It is found that the number of clouds declines with distance
from the Galactic center, with an exponential scale length $2.8 - 8\
\rm kpc$ at the 99\% confidence level. We find a lower limit to the
scale height of the clouds of 180 pc (HWHM), but the true value is
likely significantly higher.

\end{abstract}


\section{Introduction}

A large fraction of the neutral halo in the inner Galaxy is in the
form of clouds, observed in \HI\ emission to at least $|z|=1.5\ \rm
kpc$ from the Galactic plane, and downward to within $|z| \approx 200$
pc, where they become heavily blended with disk \HI\
\citep{lockman2002}.  A median halo cloud from a sample of 40 located
at a median distance from the plane of $z = -940$ pc has an \HI\ mass
of $50\ M_\odot$, a diameter of a few tens of parsecs, a column
density of a few times $10^{19}\ \rm cm^{-2}$, and average \HI\
density of a few tenths $\rm cm^{-3}$. The clouds in the halo follow
Galactic rotation, but with a fairly large cloud-to-cloud velocity
dispersion of several tens of $\kms$ \citep{lockman2002,
lockman2003b}.

It is possible that the halo clouds are part of a widespread
population, which may include the `fast' clouds seen in optical and
\HI\ absorption surveys (e.g., M\"unch \& Zirin 1961, Radhakrishnan \&
Srinivasan 1980, Mohan et~al. 2004, Welsh, Sallmen \& Lallement 2004,
Dwarakanath 2004) at high galactic latitude, and, when unresolved, as
broad components in \HI\ emission lines
\citep{kulkarni1985,kalberla98}, but see \cite{dwarakanath2004b} for a
different interpretation of broad absorption features toward SGR
A*. \citet{kulkarni1985} constrained the the mass of a population of
fast-moving clouds with a large scale height to at most 20\% of the
Galactic \HI\ mass, but noted that it could contain most of the
kinetic energy in the interstellar medium.

The clouds, however, may not be confined to the halo:
\citet{lockman2003a} reported the discovery of a small \HI\ cloud in
the VLA Galactic Plane Survey (VGPS) \citep{taylor2002} which lies
only 22 pc from $b = 0^\circ$, but which shares many characteristics
of the halo clouds.  They speculated that it might be part of the halo
population, visible at low latitude only because its random velocity
puts it more than $40\ \kms$ beyond the bulk of Galactic \HI. The VGPS
is unrivaled for study of low latitude Galactic \HI\ in the first
longitude quadrant.  In this paper we present the results of a
systematic search for clouds with velocities beyond that permitted by
normal Galactic rotation at the low Galactic latitudes covered by this
new survey.

\section{Data Set and Search parameters}

The VLA Galactic Plane Survey \citep{taylor2002} combines observations
of the 21 cm \HI\ line and continuum emission over Galactic longitudes
$18\arcdeg \leq l \leq 67^\circ$ made with the D configuration of the
Very Large Array (VLA) and the Green Bank Telescope (GBT) of the
NRAO\footnote{The National Radio Astronomy Observatory is operated by
Associated Universities, Inc., under a cooperative agreement with the
National Science Foundation.}.  The survey covers Galactic latitudes
$\pm1\fdg3$ at low longitudes to $\pm2\fdg3$ at the higher longitudes
\citep{taylor2002}.  The resolution of the VGPS is $1' \times 1'
\times 1.56\ \kms$ (FWHM) in the 21cm line and spectra extend over 240
$\kms$ at the full sensitivity of $2\ \rm K$ (r.m.s.) per channel.
This is sufficient to reach velocities well beyond that of the \HI\
emission at the higher longitudes.  Higher velocities were covered by
only one receiver channel in the VLA data.  These data were also
imaged, extending the velocity range by $\sim 30\ \kms$, although at a
sensitivity degraded by a factor $\sqrt 2$.

In the inner Galaxy the maximum permitted rotational velocity at any
longitude can be established from observations of \HI\ or molecular
clouds (e.g. Burton \& Gordon 1978, Clemens 1985) In general there is
a well-defined terminal velocity at any $|l| < 90\arcdeg$: for a flat
rotation curve $V_{\rm tan} = V_{\Sun} ( 1 - \sin l)$.  Random motions
can carry individual clouds to `forbidden' velocities $V_{\rm LSR} >
V_{\rm tan}$.  For example, some halo clouds have random velocities in
excess of $40\ \kms$ \citep{lockman2002,lockman2003b}.

Clouds at positive forbidden velocities in the VGPS were identified by
visual inspection of the data both at the full angular resolution, and
also smoothed to $2'$ resolution to improve the sensitivity to
resolved clouds of low brightness.  The data were also searched by
applying a spectral filter.  Beyond velocity $V_{\rm 2K}$, defined as
the velocity where $T_{\rm b} = 2\ \rm K$ on the positive velocity
wing of the \HI\ profiles, a search was done for occurrences of 5
consecutive channels with brightness temperature exceeding 3 times the
r.m.s. noise. The velocities searched correspond to $V_{\rm LSR} >
V_{\rm 2 K} \approx V_{\rm tan} + 25\ \kms$. The images resulting from
the spectral filtering were inspected for cloud candidates missed in
the visual examination of the data.  The final sample of clouds is
believed to be complete for velocities exceeding $V_{\rm 2 K}$ and
peak \HI\ brightness temperatures $\gtrsim 10\ \rm K$. Lines-of-sight
directly toward very bright continuum sources were not included in the
search because of the possibility of systematic effects in the \HI\
spectra.

Some cloud candidates detected in the VLA data were confirmed using
the VGPS zero-spacing dataset from the Green Bank Telescope (GBT) and
also by analyzing subsets of the VLA data taken over several different
days.  Fifteen cloud candidates were subsequently re-observed with the
GBT in October 2004 during which \HI\ spectra were taken at
integration times up to one minute in the direction of the clouds and
at $\pm 10\arcmin$ offsets in $l$ and $b$.  These deeper GBT spectra
have typical rms noise levels of $< 60\ \rm mK$.

\section{Cloud Properties}

Table~1 lists observed properties of the forbidden velocity \HI\
clouds which we have identified.  The velocity, peak brightness
temperature and line width $\Delta v$ (FWHM) were measured by fitting
a Gaussian to the line profile through the brightest part of the
cloud.  The errors listed for these quantities are $1\sigma$ errors
from the Gaussian fits.  The peak \HI\ column density, $N_{\rm HI}$,
was derived assuming that the lines are optically thin, a good
assumption unless their excitation temperature is $< 100$ K (see below
for limits on the temperature).  Cloud diameters were determined by
fitting an ellipse to the half peak $N_{\rm HI}$ contour.  The
location of the clouds in longitude and velocity is shown in
Figure~\ref{lvmap-fig}. The upper boundary of the velocity range
searched is indicated by a dashed line.  Detected clouds fall in a
band in Figure~\ref{lvmap-fig} which has a width of $\sim 60\ \kms$
and follows the variation of the terminal velocity with longitude. The
upper boundary of the velocity range of the ensemble of clouds does
not appear to be limited by the velocity coverage of the survey. The
fact that the velocities of the clouds change systematically with
longitude following the main \HI\ emission indicates that these
clouds, like \HI\ clouds in the lower halo, have kinematics dominated
by normal Galactic rotation despite their large random velocities
\citep{lockman2002}. These clouds should be considered as a disk
population.

The peak brightness temperatures of clouds in the sample lie in the
range $10$ to $22\ \rm K$. This range is limited by sensitivity on the
low end, and possibly by resolution on the high end, as some clouds in
the sample have angular diameters comparable to the $1'$ resolution.

\begin{figure}
\centerline{\resizebox{13cm}{!}{\includegraphics[angle=0]{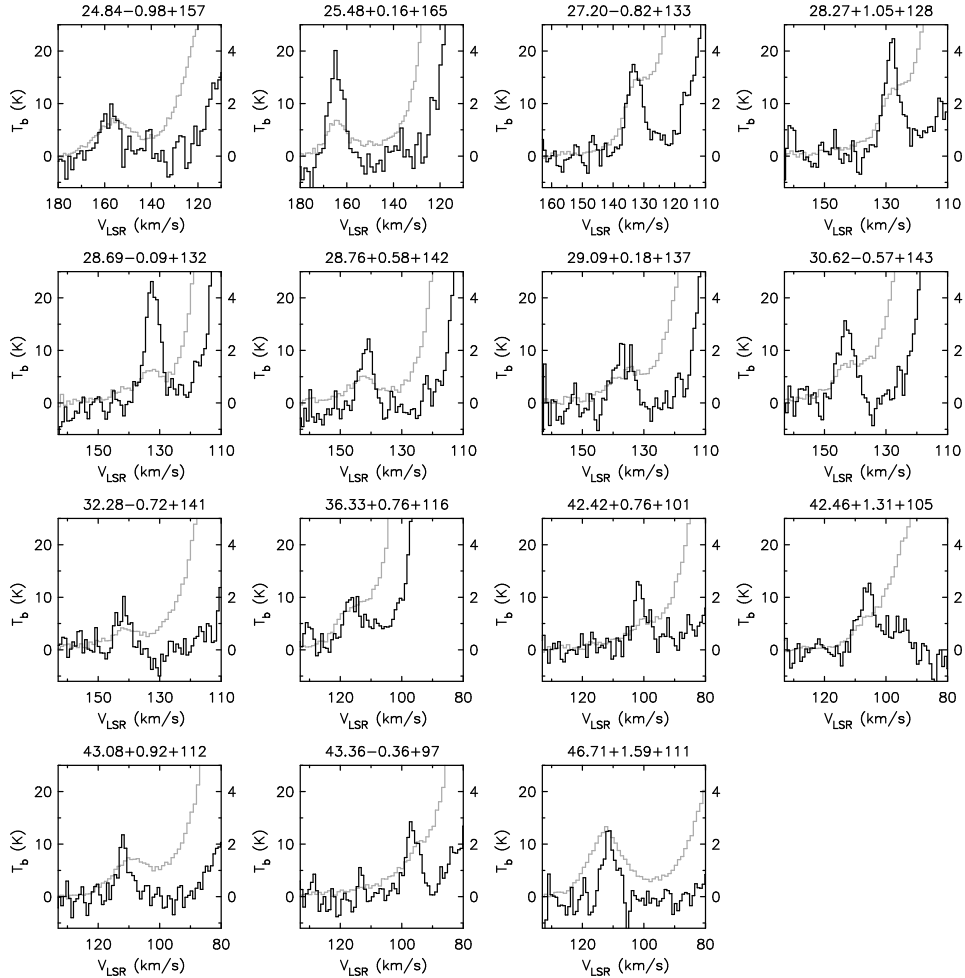}}}
\caption{ \HI\ line profiles through the center of the clouds listed
in Table~\ref{observed-tab}.  Dark lines are from VGPS data
(temperature scale on the left axis), light lines from GBT data only
(temperature scale on the right axis). }
\label{profile-fig}
\end{figure}

\begin{figure}
\centerline{\resizebox{13cm}{!}{\includegraphics[angle=0]{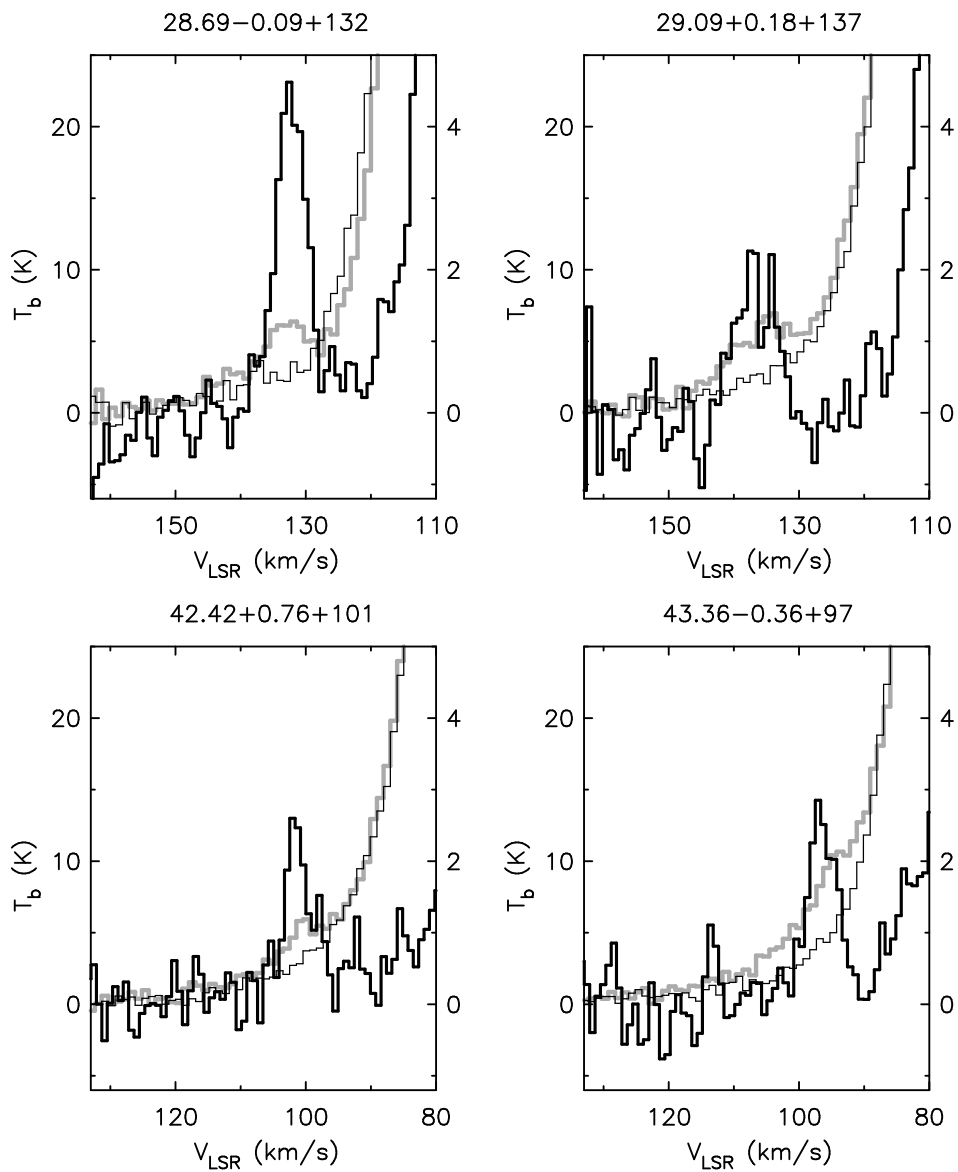}}}
\caption{ Enlargement of the line profiles of four clouds from
Figure~\ref{profile-fig}. The thick black histogram is the VGPS
profile, the thick gray histogram is the GBT profile as in
Figure~\ref{profile-fig}.  The thin black histogram shows the mean
GBT profile of the off-cloud positions. The labels on the axes are the
same as in Figure~\ref{profile-fig}.  }
\label{faintprofile-fig}
\end{figure}

The sensitive GBT spectra provided independent confirmation of the
clouds in the sample.  \HI\ line profiles from the VGPS are shown in
Figure~\ref{profile-fig} with the \HI\ profiles from the follow-up
observations made with the GBT.  Figure~\ref{faintprofile-fig} shows
enlargements of the profiles of four clouds which are difficult to
discern Figure~\ref{profile-fig}, in conjunction with the mean of the
profiles in the off-cloud positions. A significant excess over the mean
off-cloud profile at the velocity of the VGPS cloud is seen in all
profiles except for the cloud $42.46+1.31+105$ which suffers from
inconsistencies between the profiles at the offset positions. The
shoulder in the GBT profile of this cloud provides a tentative
detection in the GBT data.

The median size of clouds in the present sample is $3\farcm4$,
resulting in a significantly lower peak brightness temperature in the
GBT spectra because of beam dilution.  The peak brightness temperature
in the GBT profiles is less than 3 K, which is much smaller than the
$\sim 15\ \rm K$ peak brightness temperature at $1'$ resolution
through the brightest part of the clouds.  The VGPS data suggest that
values of $N_{\rm HI}$ for these clouds derived from GBT data alone
would be underestimated by a factor $\sim 7$ in the median, and values
of $\langle n_{\rm HI}\rangle$ by a factor $\sim 20$.  A clear example
is $25.48+0.16+165$.  The peak brightness temperature in the GBT
profile is a factor 10 smaller than in the VGPS profile, consistent
with beam dilution of the $3'$ diameter cloud.

The mean GBT spectra of the off positions were also inspected for
evidence of an extended ($>10'$), low-column density component with
the same velocity as the clouds identified in the VGPS. The
sensitivity of this experiment is determined by the presence of a
faint wing of the Galactic line profile which appears as a
continuation of the line profile of emission at lower, permitted,
velocities as shown in Figure~\ref{profile-fig} and
Figure~\ref{faintprofile-fig}. No evidence for an extended component
with the same velocity as the cloud was found. In the median, the peak
brightness temperature of a cloud in the VGPS data is $\sim 20$ times
brighter than the wing of the Galactic \HI\ profile at the velocity of
the cloud as determined from the average GBT spectra in the off
positions. This indicates a large contrast between the clouds detected
in the VGPS and any extended \HI\ component with the same
line-of-sight velocity.  It should be noted that clouds with a large
brightness contrast are easier to detect, so this number does not
necessarily apply to the general population of fast-moving \HI\
clouds.

\HI\ column density maps were derived by summing over channels within
$\pm 1.25 \Delta v$ (three times the Gaussian velocity dispersion) of
the central velocity of the cloud.  Figure~\ref{NHImap-fig} shows a
selection of column density maps representative for the sample. Many
clouds are compact, isolated \HI\ features, such as $25.48+0.16+165$
and $43.08+0.92+112$. For the latter, there is some detached emission
$0\fdg25$ from the brightest part, which is not listed as a separate
cloud.  The cloud $46.71+1.59+111$ is well resolved into a structure
with multiple compact components.  The most complicated structure is
seen near longitude $29\degr$ (upper right panel in
Figure~\ref{NHImap-fig}). The emission is distributed over $\sim 1
\degr$.  The emission near the center of the image was listed as
$28.76+0.58+142$. The emission in the lower part of this panel is part
of $29.09+0.18+137$ and $28.69-0.09+132$.  The willing eye could
divide the structure seen in this part of the survey into more clouds
than those listed in Table~\ref{observed-tab}. This was not done
because we aim to avoid counting substructure as individual clouds.
This issue will be discussed in more detail in Section~\ref{select-sec}.

Table~2 lists derived parameters of the clouds in the sample.  The
distance adopted is $d = R_0 \cos l$ with $R_0 = 8.5\ \rm kpc$, which
assumes that each cloud is at the tangent point. This is a good
assumption on average, but the uncertainty in the distance of an
individual cloud can be large. A discussion of the error in the
distance is deferred to Section~\ref{model-sec}. A lower limit to the
random component of the velocity of the clouds is given by the
parameter $V_{\rm pec} \equiv V_{\rm LSR} - V_{\rm tan}$ where the
terminal velocity, $V_{\rm tan}$, is derived from the observations of
$\rm ^{12}CO$ emission in the plane by \citet{clemens}.  This
definition of $V_{\rm tan}$ takes advantage of the small random
velocities of molecular clouds to infer Galactic rotation, while
including information on the regular variations observed in the
tangent-point rotation curve.  The effect on this analysis of possible
non-circular motions was tested by examining the dynamical simulations
of \citet{weiner}, which include the effects of perturbations caused
by a Galactic bar.  Over the longitudes considered here the highest
$V_{\rm LSR}$ occurs typically within $5\%$ of the distance to the
tangent point, and the largest displacement is only $20\%$.  It
appears that the identification of the highest velocity in a \HI\
spectrum as arising from the geometric tangent point is quite robust
in the first galactic quadrant outside the galactic center.  The
derived values of $V_{\rm pec}$ are lower limits in the case that the
cloud is at a distance other than the tangent point.

Cloud dimensions were calculated from the angular sizes in Table~1
deconvolved with the $1'$ VGPS beam assuming a Gaussian shape.  The
deconvolved angular size was larger than $1'$ for all clouds. Formally
all clouds in the sample are resolved by the $1'$ beam, but some
clouds are only marginally resolved. The \HI\ number density $\langle
n_{\rm HI} \rangle$ (Table~\ref{derived-tab}) is defined as the peak
column density divided by the harmonic mean of the major and minor
dimensions of the cloud.  The line width of the cloud provides an
upper limit to the temperature of a cloud because the thermal width
cannot exceed the observed width of the line profile. This leads to
the upper limit $T_{\rm kin} \leq 22 \Delta v^2$, with $T_{\rm kin}$
in K and $\Delta v$ the FWHM width of the line profile in $\kms$. The
density and limits to the temperature in Table~\ref{derived-tab} refer
to a line-of-sight through the brightest part of the cloud.  For the
larger clouds $59.67-0.39+60$, and $60.70+1.02+58$, average values for
several positions across the clouds were used.

\section{Selection Effects}
\label{select-sec}

\subsection{Velocity range}
\label{select_vel_sec}

\begin{figure}
\centerline{\resizebox{13cm}{!}{\includegraphics[angle=0]{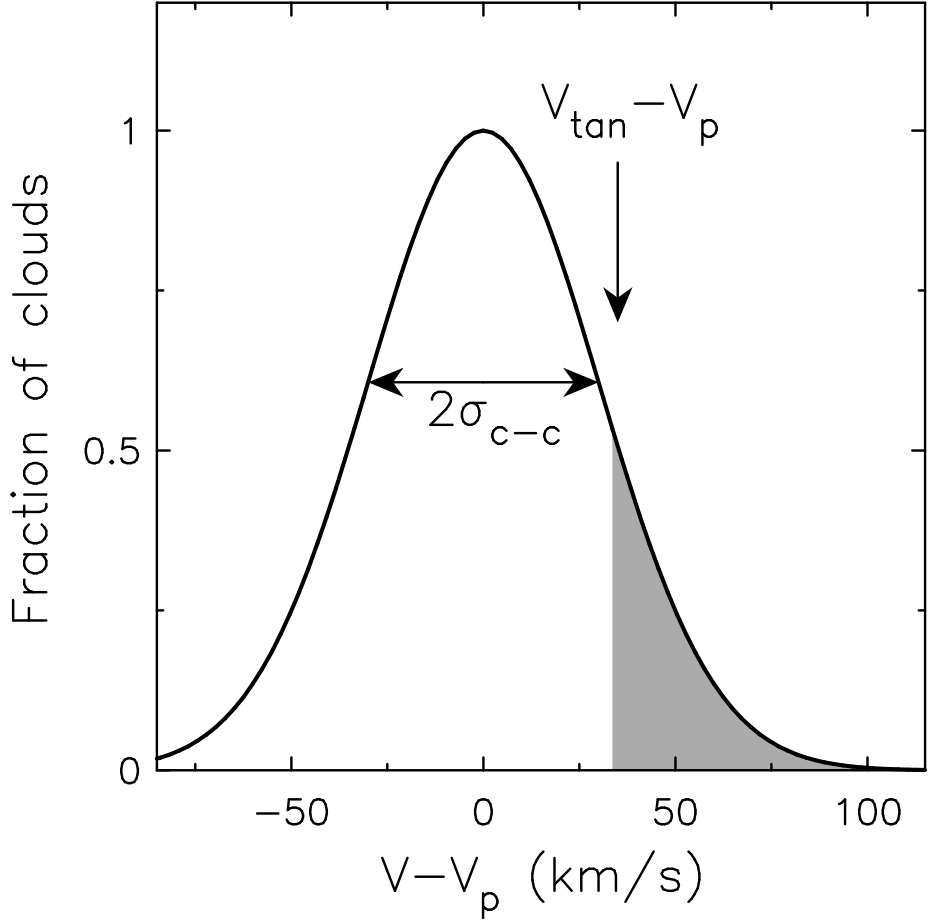}}}
\caption{ The line-of-sight velocity distribution of clouds at
locations along the line-of-sight which correspond to a velocity
$V_{\rm p}$ according to Galactic rotation.  The shaded area indicates
the fraction of clouds oberserved at velocities $V > V_{\rm
tan}$. This diagram illustrates the case for $l=27\degr$,
$\sigma_{c-c} = 30\ \kms$, and for clouds at distances 5.1 and 10.1
kpc from the Sun.
\label{veldisp-fig}
}
\end{figure}

Clouds may occur anywhere along the line-of-sight, but only those
clouds with large forbidden velocities are counted. If a cloud is
located in distance far from the tangent point, it must have a large
random velocity component along the line-of-sight to be observed at a
forbidden velocity. A cloud near the tangent point may have a smaller
random velocity component along the line-of-sight in order to be
observed at the same forbidden velocity, as illustrated in
Figure~\ref{veldisp-fig}. Consider a velocity $V_{\rm p}$ with $V_{\rm
p} < V_{\rm tan}$. In the first Galactic quadrant the velocity $V_{\rm
p}$ will arise from more than one location along the line-of-sight.
Clouds at those locations have a line-of-sight velocity distribution
which is represented by a Gaussian in Figure~\ref{veldisp-fig}. Some
of these clouds appear at velocities larger than the tangent point
velocity $V_{\rm tan}$ as indicated by the shaded area in
Figure~\ref{veldisp-fig}. If $V_{\rm p}$ is much smaller than $V_{\rm
tan}$, only a small fraction of the clouds at the locations
corresponding to $V_{\rm p}$ appear at forbidden velocities. This
creates a bias for clouds observed at forbidden velocities to be
located around the tangent point. The strength of this bias depends on
the kinematic properties of the cloud population and on the direction
of the line-of-sight. The total number of clouds at forbidden
velocities is found by integration over all $V_{\rm p}$.

The velocity range of the search area is limited on the high-velocity
side by the edge of the spectral band of the data, and on the
low-velocity side by Galactic \HI\ emission at permitted velocities.
The high end of the velocity range extends far beyond the terminal
velocity of Galactic \HI.  The velocity range of the search was
extended to include the area sampled only by the high-velocity
receiver channel of the VGPS VLA data.  The absence of clouds detected
near the upper velocity boundary of the survey
(Figure~\ref{lvmap-fig}) indicates that the velocity coverage of the
survey does not introduce a significant selection effect in the number
of clouds detected.  It is found that the number of clouds decreases
rapidly as the velocity increases beyond the tangent point velocity.

On the low-velocity side, a limit must be set to identify clouds with
a large velocity beyond the terminal velocity.  The actual maximum
permitted velocity depends in a complicated way on Galactic longitude
because of non-circular orbits of the gas and streaming motions
associated with spiral arms. The effect of a spiral arm on the maximum
permitted velocity becomes important if the spiral arm is close to the
tangent point. A relatively large number of \HII\ regions and
supernova remnants associated with the spiral arm may be located close
to the terminal velocity at the longitude of a spiral arm
tangent. Disturbances of the interstellar medium by these objects are
more likely to produce emission at forbidden velocities at these
longitudes.  Various tracers of the spiral arm tangents compiled by
\citet{englmaier1999} show that within the VGPS area, the
line-of-sight becomes a tangent to the Sagittarius arm at $l \approx
51\degr$ and to the Scutum arm at $l = 24\degr$ to $l = 30\degr$. We
note that half of the sample of clouds in Figure~\ref{lvmap-fig}
appears in the longitude range of the Scutum arm tangent. However, the
remaining half of the sample does not appear near a spiral arm
tangent.

The lower velocity limit of the cloud search ($V > V_{\rm 2 K}$) is
defined by the data, not by a model for Galactic rotation. This
definition of the lower velocity limit of the search area takes into
account the variations of the terminal velocity associated with spiral
arms, and thus ensures homogeneous selection criteria for all
longitudes.  The lower velocity limit of the search area can be
expressed in approximation in terms of $V_{\rm tan}$ by noting that
for longitudes larger than $\sim 30^\circ$, $V_{\rm 2 K} \approx
V_{\rm tan} + 25 \ \kms$. The peculiar velocity along the
line-of-sight, $V_{\rm pec}$, was determined relative to the maximum
velocity of molecular clouds at the same longitude, which is
determined more accurately than the \HI\ terminal velocity.

\subsection{Survey area and geometric selection effects}

The VGPS survey area covers the first Galactic quadrant from $l =
18^\circ$ to $l = 67^\circ$. The latitude coverage of the VGPS is at
least $\pm 1\fdg3$, but increases to $\pm1\fdg9$ from longitude
$46^\circ$ to $59^\circ$, and to $\pm 2\fdg3$ from longitude
$59^\circ$ to $67^\circ$. An outline of the survey area was given by
\citet{taylor2002} and \citet{lockman2003a}. The varying latitude
coverage of the survey must be taken into account in the analysis of
the Galactic distribution of the clouds. The importance of the
selection effect of the varying latitude coverage depends in part on
the distribution of the clouds in the Galaxy. Clouds in the sample
tend to be located near the tangent point
(Section~\ref{select_vel_sec}). The mean distance and Galactocentric
radius of clouds in the sample vary accordingly with longitude. The
larger distance of the tangent point at low longitudes compensates for
the smaller latitude coverage of the survey, as illustrated by the
following example.  For a constant cloud scale height, at $l=25\degr$,
the maximum vertical distance at the tangent point probed by the
search for clouds is $\pm 182\ \rm pc$, while at $l=60\degr$ it is
$\pm 156\ \rm pc$. If the scale height of the cloud population is much
less than 100 pc, most of the clouds will be located within the survey
area at all longitudes. The effect of the varying latitude coverage of
the survey would be small in this case.  However, if the scale height
of the cloud population is much more than 200 pc, the fraction of the
scale height probed by the VGPS is larger at low longitudes than at
high longitudes. Thus, for a large cloud scale height, a constant
number of clouds per unit volume would produce {\it fewer} clouds in
the VGPS at high longitude, despite the larger latitude coverage
there.

\begin{figure}
\centerline{\resizebox{13cm}{!}{\includegraphics[angle=0]{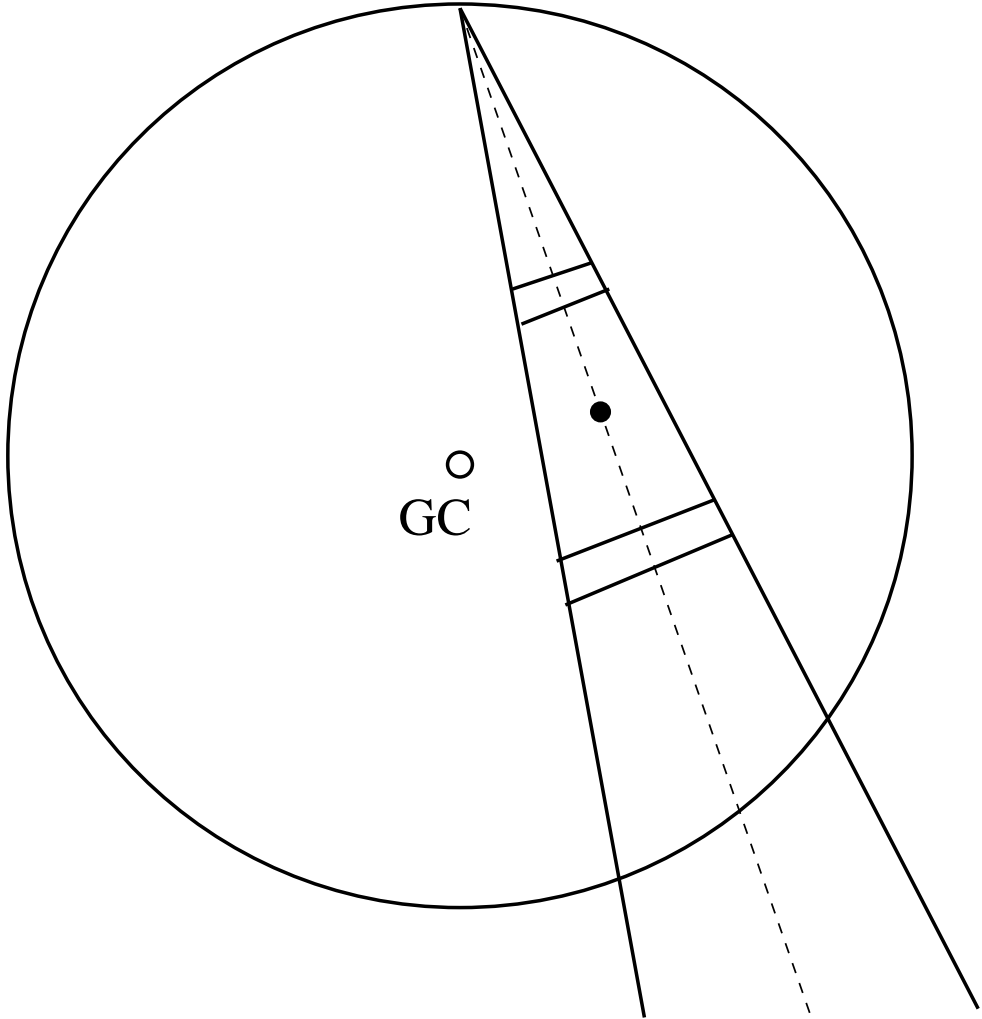}}}
\caption{ Illustration of the asymmetry of the volume probed at the
near side and the far side of the tangent point, affecting the number
of clouds observed at forbidden velocities. The circle indicates the
orbit of the sun around the Galactic center (GC). The filled circle is
the tangent point for the line-of-sight indicated by the dashed line.
The near and far side volume elements shown here approximately
represent locations with a velocity $30\ \kms$ below the tangent point
velocity.
\label{los-fig}
}
\end{figure}

This argument assumes that all clouds are located at the tangent
point.  In fact, clouds detected at forbidden velocities occur along a
range of distances along the line-of-sight, depending on the
cloud-cloud velocity dispersion. Each positive velocity can arise from
two locations (Figure~\ref{los-fig}).  If the cloud-cloud velocity
dispersion is large, the difference in distance between the near and
far locations can be significant, and the volume probed at the far
side is correspondingly larger. At higher longitudes ($l \gtrsim
50\degr$), clouds in the outer Galaxy may become visible at positive
forbidden velocities. These clouds are always located behind the
tangent point. In this case, the assumption that the distance of a
cloud is on average the distance to the tangent point, breaks
down. However the error introduced by assuming a tangent point
distance remains negligible compared with the uncertainty in the
distance of an individual cloud associated with the large random
component of the line-of-sight velocity.

Interpretation of the relative numbers of clouds seen at high and low
longitude requires careful modeling of the geometric selection effects
related to the survey area and the distribution and kinematics of the
ensemble of fast-moving clouds. Such models will be presented 
in Section~\ref{model-sec}.

\subsection{Identification of clouds}

Reliably identifying compact, faint emission in a large dataset such
as the VGPS requires limits to avoid spurious detections from random
fluctuations of the noise.  Candidate detections were required to have
a peak brightness temperature more than $10\ \rm K$, and 5 consecutive
channels above $6\ \rm K$ (3 times the r.m.s. noise per channel). 
Cloud candidates were confirmed by follow-up observations with the
GBT, except for the two larger clouds near longitude
$60^\circ$. The sample presented here is believed to be complete
within the specified limits of the search.

Several clouds in the sample appear as compact, isolated objects in
the VGPS images.  Other clouds are resolved, and may appear as more
than one component. In the present search, multiple features with
nearly the same velocity (compared with the line width of the
brightest component) are considered to be a part of the same
cloud. This approach aims to avoid a subjective subdivision of related
emission into ``sub-clouds'', which could bias the analysis of the
Galactic distribution of these clouds addressed in this paper.

Occasionally, \HI\ emission is found which appears as a localized
broad wing of the Galactic \HI\ profile. Two approximately circular
areas of $\sim 1\degr$ in diameter were identified in the VGPS data,
and are shown in Figure~\ref{group-fig}a,b.  The structure at
longitude $l = 25\degr$ is also visible in data from the
Leiden-Dwingeloo survey \citep{hartmann} shown by
\citet{koo2004}. These structures are resolved into a number of
smaller components in the arcminute-resolution VGPS images. Although
some of the substructure would qualify as a ``fast-moving cloud''
according to our selection criteria, these structures were not
included in the analysis of this paper. This is because these larger
structures appear as wings on the Galactic line profile, not as
objects with a distinct, well-defined forbidden velocity.  The feature
at $l \approx 25\degr$ has a mass $M_{\rm HI} = 5\ \pm\ 2\ \times 10^4
\ \rm M_\sun$, assuming a tangent point distance. The mass of the
feature at $l \approx 31\degr$ is $M_{\rm HI} = 4\ \pm\ 1\ \times 10^4
\ \rm M_\sun$ assuming a tangent point distance. The main uncertainty
in the mass determination is confusion with emission at permitted
velocities.  Broad wings of the Galactic \HI\ line may be associated
with supernova remnants or stellar wind bubbles with a large expansion 
velocity \citep{koo1990,koo1991}.  The structures in Figure~\ref{group-fig}a,b
do not have a counterpart in the VGPS continuum images, but this does
not exclude the possibility that these wings originate from old
supernova remnants which are no longer visible in the continuum
\citep{koo2004}.

In this paper we concentrate on the small isolated clouds seen at
forbidden velocities, which may or may not be related to the broad
\HI\ wings associated with old supernova remnants. We note that the
cloud $25.48-0.16+165$ appears in the direction of the larger
structure at $l \approx 25\degr$. It is listed as an isolated cloud
because of its much larger velocity, but it is possible that this
cloud is associated with the larger structure.

\section{Galactic distribution of the clouds}
\label{model-sec}

\begin{figure}
\centerline{\resizebox{13cm}{!}{\includegraphics[angle=0]{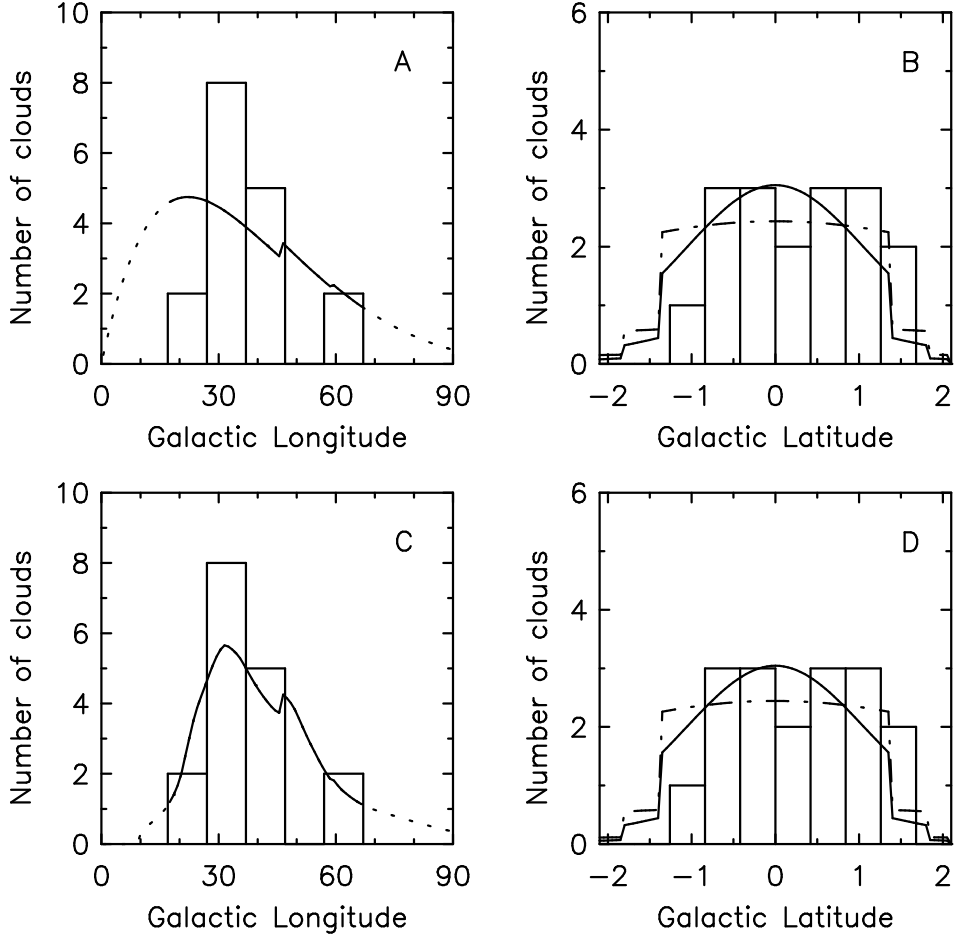}}}
\caption{ Distribution in longitude and latitude of the clouds
compared with models for the spatial distribution of the
clouds. Dotted curves show the model predictions outside the VGPS
survey area. These parts were not included in the Kolmogorov-Smirnov
tests. The models were scaled so that area under the model curves
within the VGPS survey limits is the same as the area under the
histograms.  A: Model of class 1 in longitude with radial scale length
4.5 kpc, vertical scale height 180 pc, and $\sigma_{\rm c-c} = 30\
\kms$.  B: Latitude distribution of clouds compared with models of
class 1 with vertical scale height 180 pc (solid line) and 540 pc
(dot-dashed line).  C: Model of class 2 in longitude, with scale
height and $\sigma_{\rm c-c}$ as in A. D: Latitude distribution of
clouds compared with models of class 1 with vertical scale height 180
pc (solid line) and 540 pc (dot-dashed line).
\label{clouddist}
}
\end{figure}

The distribution of the clouds in longitude and latitude is shown in
Figure~\ref{clouddist}. The longitude distribution is most sensitive
to a variation of the number of clouds with distance from the Galactic
center. The latitude distribution is most sensitive to the scale
height of the ensemble of clouds.  The observed distributions are also
affected by observational selection effects discussed in
Section~\ref{select-sec}.

The longitude distribution in Figure~\ref{clouddist} peaks near
longitude $\sim 30^\circ$.  The latitude distribution of the clouds is
nearly flat. To use these data to constrain the Galactic distribution
of the ensemble of clouds, the observed distributions were compared
with simple models of the space distribution of the clouds integrated
along the line-of-sight, including observational selection effects.

The Galactic distribution of the clouds is modeled as a layer with a
surface density (number of clouds kpc$^{-2}$) which may vary with
Galactocentric radius, and a scale height and cloud-cloud velocity
dispersion which do not vary with position. The scale height and
velocity dispersion are treated as independent parameters because we
do not wish to impose an assumption of hydrostatic equilibrium on the
models by coupling the cloud-cloud velocity dispersion to a scale
height.

Only a fraction of a population of fast moving clouds along the
line-of-sight will appear at forbidden velocities. This fraction
depends on the kinematic properties of the ensemble of clouds, and on
the direction of the line-of-sight. The random line-of-sight velocity
components of the clouds are expressed in the form of a probability
distribution $f(v)$ with a single cloud-cloud velocity
dispersion. Assuming an axially symmetric distribution of clouds
$n_c(R)$ and a flat Galactic rotation curve ($R_0 = 8.5\ \rm kpc$,$V_0
= 220\ \kms$) with circular orbits, the number density of clouds at
velocity $V_{\rm p}$ can be expressed as a function of velocity
$n_c(V_{\rm p})$.  Let $\Phi(V_{\rm p})$ be the volume of space which
corresponds with the velocity interval $\langle V_{\rm p},V_{\rm
p}+dV_{\rm p} \rangle$.  $\Phi(V_{\rm p})$ is a complicated function
which contains details of the latitude coverage and geometrical
selection effects.  With these assumptions, the number of clouds with
a line-of-sight velocity exceeding a limiting velocity $V_{\rm lim}$
at a particular longitude $l$, is
$$
N_c(l)=\int_{-\infty}^{V_{\rm tan}}  \ n_{\rm c}(V_{\rm p})\ \Phi(V_{\rm p})\ \bigg\{ \int_{V_{\rm lim}}^{\infty} f(v-V_{\rm p}) dv \bigg\}\ d V_{\rm p} \eqno(1)
$$ The velocity distribution of the population of clouds thus acts as
a weighting function in the integral of the number of clouds along the
line-of-sight. The lower limit of the outer integral should take into
account that at higher longitudes in the VGPS survey area clouds
located outside the solar circle ($R > R_0$) may in principle be seen
at positive forbidden velocities.  Little is known about the shape of
the velocity distribution $f(v)$; we assume it is a Gaussian. The
effect of replacing $f(v)$ with a function with non-gaussian wings
\citep{siluk1974} is negligible for the current data.

The sample was selected with $V_{\rm lim} = V_{\rm 2K} \approx V_{\rm
tan} + 25\ \kms$. Inserting $V_{\rm lim} = V_{\rm tan}$ changes the
normalization of the distribution $N_c(l)$ significantly, and hence
the total number of clouds, but the shape of the distribution remains
virtually the same for the longitude range covered by the VGPS.

The value of the cloud-cloud velocity dispersion is not well
constrained by the data.  The number of clouds at the tangent point
velocity cannot be determined because of confusion with emission at
permitted velocities. However, the distribution of the clouds in
longitude and velocity can provide a plausible range for the value of
the velocity dispersion. Velocity crowding at the tangent point
velocity projects a large faction of the line-of-sight into a small
velocity range.  As a first approximation, each cloud can be assumed
to originate from a location with (nearly) the tangent point velocity.
If the cloud-cloud velocity dispersion were significantly larger than
$50\ \kms$, more clouds with peculiar motions in excess of $50\ \kms$
would have been found.  Also, the broad distribution of clouds in
velocity compared with the wing of emission at permitted velocities
suggests that the cloud-cloud velocity dispersion is larger than $\sim
10\ \kms$.  The range of the cloud-cloud velocity dispersion adopted
in the sequence of models is $20\ \kms$ to $50\ \kms$ in steps of $10\
\kms$. This range covers the velocities of clouds in the solar
neighbourhood seen in optical and \HI\ absorption experiments.

The primary effect of a larger cloud-cloud velocity dispersion is to
make a larger fraction of the ensemble of clouds visible at forbidden
velocities.  If the model distributions are normalized to the observed
distribution, the number density of clouds $\rm kpc^{-3}$, $n_c$, and
the cloud-cloud velocity dispersion, $\sigma_{\rm c-c}$, are
correlated to make the product $n_c \sigma_{\rm c-c}^{0.5}$
approximately constant (see Appendix~A). This is the quantity which is
constrained by the data.  Models in Figure~\ref{clouddist} assume a
cloud-cloud velocity dispersion $\sigma_{\rm c-c} = 30\ \kms$. Models
with a different value for $\sigma_{\rm c-c}$ would give nearly
identical scaled distributions, but imply a different number density
$n_c$.

Two prescriptions for the variation of the surface density of clouds
in the disk were considered. Class 1 models use an exponential
function of Galactocentric radius with the radial exponential scale
length as a free parameter. Class 2 models assume that the surface
density of clouds follows the surface density of molecular gas in the
Galactic disk according to \citet{dame1993}.  A class 1 model has
three input parameters (scale length, scale height and cloud-cloud
velocity dispersion). A class 2 model has two input parameters (scale
height and cloud-cloud velocity dispersion). The total number of
clouds is also a free parameter because of the scaling of the models
to the observed distributions.  The vertical distribution of clouds is
represented by a Gaussian. The vertical scale height is defined as the
location where the number of clouds drops to half the number in the
mid plane.  The models were integrated along the line-of-sight within
the VGPS survey area, and integrated in latitude and longitude to
obtain model longitude and latitude distributions.

A Kolmogorov-Smirnov (KS) test was used to test the significance of
differences between the observed distributions and the models.  A
large number of models was created, representing wide ranges in the
radial scale length, vertical scale height, and cloud-cloud velocity
dispersion.  Models were tested against the observed cloud
distributions and rejected at the 99\% confidence level.  The KS tests
were done separately for the longitude and the latitude
distributions. Class 1 models failed the KS test outside the range of
radial scale-lengths 2.8 kpc to $\sim$8 kpc. The high end of this
range is more sensitive to the assumed scale height and cloud-cloud
velocity dispersion, because a large scale height and a large
cloud-cloud velocity dispersion have a more noticeable effect at the
high longitude end of the VGPS.  Among class 1 models, those with a
radial scale length of $\sim 4.5\ \rm kpc$ fit the data best. None of
the models of class 2 was rejected by the KS tests. The small size of
the sample does not allow us to discriminate between class 1 models
and class 2 models. The correspondence between class 2 models and the
data is interesting, but not necessarily unique.

KS tests against the latitude distribution of the clouds lead to a
lower limit for the scale height of the clouds. Class 1 and class 2
models with the same scale height are nearly indistinguishable in the
predicted latitude distribution (Figure~\ref{clouddist}).  Models
assuming a Gaussian layer with scale height smaller than 180 to 200 pc
were rejected. The precise value depends somewhat on the assumed
radial scale length and the cloud-cloud velocity dispersion. Panels B
and D in Figure~\ref{clouddist} show the latitude distribution for
models with the smallest scale height not rejected by the data (180
pc), and for a scale height three times this lower limit. Models with
a large scale height fit the observed latitude distribution better.

The best-fitting model distributions can in principle be used to
estimate the number of fast-moving clouds in the Galaxy. The result of
such an estimate is very uncertain because of the number of
assumptions which has to be made. The largest uncertainties result
from the correction factor $f_b$ for the number of clouds outside the
VGPS latitude range (related to the unknown scale height of the
ensemble of clouds), and the factor $f_v$ which corrects for the
effect of the lower velocity cut-off $v > V_{\rm 2K}$ on the total
number of clouds (a strong function of the unknown
$\sigma_{c-c}$). The uncertainty in these factors is much larger than
the differences between class 1 and class 2 models. We adopt a scale
height of $540\ \rm pc$ (HWHM), which is 3 times the lower limit to
the scale height of the clouds. In this case we have $f_b = 2.7$,
which varies by a factor $\sim 3$ up or down in the range of scale
heights from $180\ \rm pc$ to $2\ \rm kpc$. Assuming $\sigma_{\rm c-c}
= 30\ \kms$, we find $f_v = 3.1$, which varies by a factor $\sim 3$ in
the range $20\ \kms < \sigma_{\rm c-c} < 50\ \kms$. The class 1 model
with radial scale length $4.5\ \rm kpc$ results in a surface density
of fast moving clouds in the solar neighborhood ($R = 8.5\ \rm kpc$)
of $\sim 7$ clouds $\rm kpc^{-2}$. The uncertainty in this number is
at least an order of magnitude because of the uncertainties in $f_b$
and $f_v$. Other factors, such as the radial scale length of the
distribution and the question of whether an exponential profile
accurately represents the Galactic distribution of the clouds,
increase the uncertainty further. This surface density of clouds
implies a mass surface density of $4 \times 10^{-4} M_\odot\rm \
pc^{-2}$ in the form of fast moving clouds in the solar neighborhood.
This is only a small fraction of the $5\ M_\odot\ \rm pc^{-2}$ surface
density of \HI\ in the Galactic disk \citep{dickeylockman}. When
comparing these numbers, it should be kept in mind that the search for
fast-moving clouds in the VGPS data is limited to clouds with fairly
high column densities. Fast-moving clouds with column densities below
the sensitivity limit of the VGPS have been observed in absorption and
in emission.  It is not clear at this time which fraction of the total
population of fast-moving clouds is detected in the VGPS.

\begin{figure}
\centerline{\resizebox{13cm}{!}{\includegraphics[angle=0]{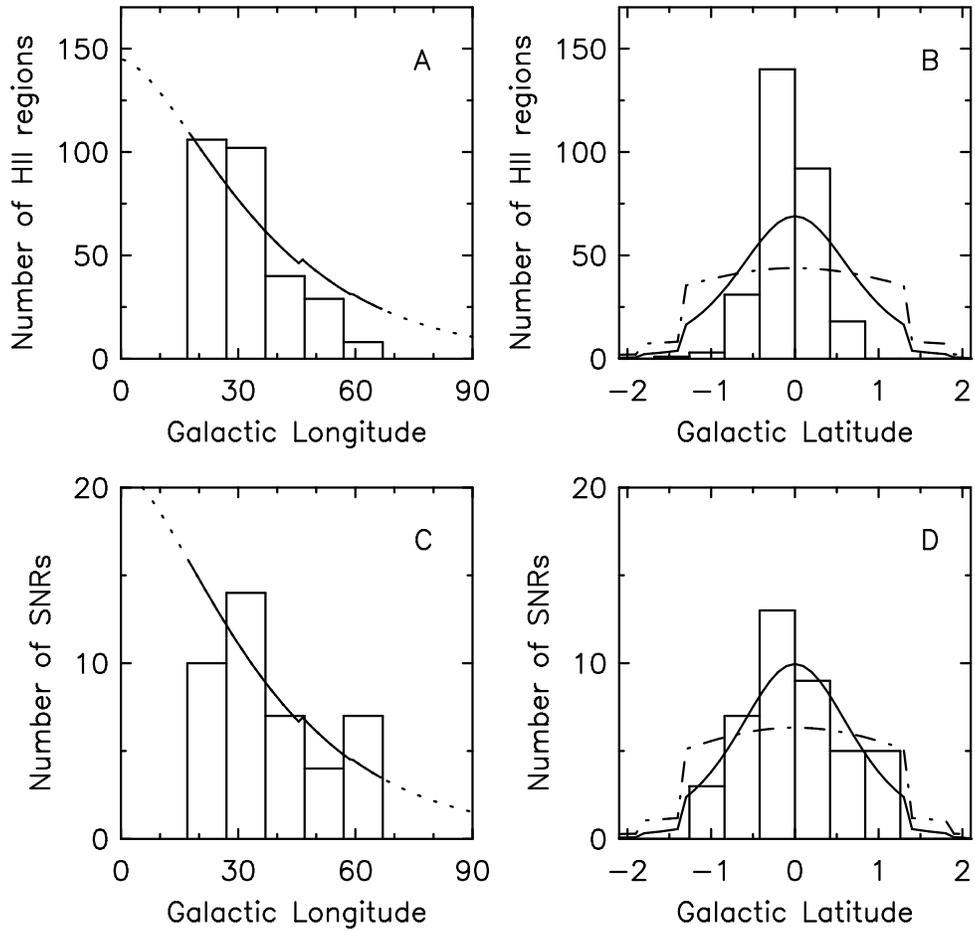}}}
\caption{ Distribution in longitude and latitude of Galactic \HII\
regions and supernova remnants (histograms) and the model of class 1
shown in Figure~\ref{clouddist}A,B. The models were integrated over
the line-of-sight without velocity weighting.
\label{HIISNR-fig}
}
\end{figure}

We also compared the best fitting model cloud distribution with the
distribution of known \HII\ regions \citep{lockman1989,lockman1996}
and supernova remnants \citep{green2004} within the VGPS survey area.
\HII\ regions and supernova remnants represent the most common places
where matter is deposited into the interstellar medium with large
velocities. The class 1 model fits shown in Figure~\ref{clouddist}A,B
were integrated along the line-of-sight and over all velocities to
create the total model longitude and latitude distributions in
Figure~\ref{HIISNR-fig}.  The distribution of \HII\ regions declines
faster with longitude than the model.  The comparison with the
supernova remnants is less clear. The model with the smallest scale
height of the clouds coincides with the observed latitude distribution
of known supernova remnants. However, it should be noted that the
lower limit to the scale height of the clouds determined here appears
to be limited mostly by the latitude coverage of the VGPS -- the true
value is likely much higher.

\section{Discussion and conclusions}

We present a sample of small neutral hydrogen clouds discovered in the
VGPS with velocities up to $\sim 60\ \kms$ beyond the local terminal
velocity.  The distribution in longitude and velocity of the clouds
(Figure~\ref{lvmap-fig}) suggests that they follow Galactic rotation
despite their large peculiar velocities. This connection with the disk
indicates that these clouds are not compact high-velocity clouds,
which do not follow Galactic rotation (e.g. De Heij et~al. 2002).
On average, the distance to the tangent point is a good estimate of
the distance for each cloud. The uncertainty in the distance of each
individual cloud is large (2 to 3 kpc for $\sigma_{\rm c-c} = 30\
\kms$), but we can characterize the median properties of clouds in the
sample.  The median \HI\ mass is $60\ M_\sun$, and the median
diameter is $8\ \rm pc$.  The median central density of a cloud is $7\
\rm cm^{-3}$.  Many clouds are not well resolved in the present data,
so the actual density may be higher than listed in
Table~\ref{derived-tab}. The clouds in the sample show a significant
spread around these median parameters.

Peculiar velocities of the clouds were determined with respect to the
terminal velocity of molecular gas. As such, the peculiar velocities
derived here are lower limits to the random components of the
line-of-sight velocity of each cloud.  If a cloud is not located at
the tangent point, the actual random component of the line-of-sight
velocity is larger.  The velocity of a cloud with respect to the
ambient medium will also have a component perpendicular to the
line-of-sight which is not constrained by the data. The largest
peculiar velocities in the sample are in the range $50$ to $60\ \kms$
(three clouds). For all clouds in the sample $V_{\rm pec}$ is a
velocity component parallel to the Galactic plane. Any mechanism which
aims to explain the origin of these clouds should produce this large
velocity component in the Galactic plane, even at very small distances
from the mid plane.

We note that the presence of small clouds with high {\it forbidden}
velocities implies that some small \HI\ clouds seen at {\it permitted}
velocities may be displaced in velocity by as much as $60\ \kms$.
Such clouds would not be noticed as peculiar, unless they appear at
velocities with little confusing emission, such as inter arm
regions. Clouds with forbidden positive velocities in the second
Galactic quadrant \citep{higgs2001,kerton2002} may be members of the
same population of fast moving clouds.  The models discussed in
Section~\ref{model-sec} show that distances to these clouds in the
second and third quadrant are very poorly constrained.

The models presented in Section~\ref{model-sec} allow an evaluation of
the Galactic distribution of the clouds without assumptions about the
distance to any individual cloud. A sequence of models was generated
to predict the sky distribution of clouds in longitude and latitude,
taking into account the selection criteria.  The hypothesis that the
observed cloud distribution could have been drawn from a particular
model distribution was tested with a KS test. The KS test uses the
cumulative distribution of observed positions, and does not depend on
the binning of the data.

The models not rejected by the KS test suggest that the number of
clouds decreases with distance from the Galactic center, and that the
scale height of the cloud population is large. Models with a scale
height less than $180\ \rm pc$ (HWHM) are rejected at the 99\%
confidence level. Much larger scale heights ($> 1\ \rm kpc$) fit the
data better.  Models which assume that the surface density of the
clouds follows the distribution of molecular gas in the Galactic disk,
or an exponential profile with a scale length of $\sim 4.5\ \rm kpc$
fit the observed longitude distribution best. The difference between
these models is not statistically significant. The radial models would
easily be distinguished by observations with the VGPS sensitivity and
resolution at longitudes less than $20\degr$. Similar observations
with a larger latitude coverage than the VGPS would be valuable to
determine the scale height of these clouds.

The number of clouds in the disk declines with distance from the
Galactic center in a similar way as tracers of massive star
formation. However, the clouds have a much larger scale height than
tracers of star formation such as known \HII\ regions in the VGPS
survey area (Figure~\ref{HIISNR-fig}). The connection of the clouds
with a large scale height and massive star formation in the disk may
be established through the galactic fountain model
\citep{bregman1980}. However, this remains speculative, as the only
connection between the clouds described here and star formation in the
disk is that both seem to have approximately the same radial
distribution. A significantly larger sample of fast moving clouds in
the disk is required to confirm this result.

The clouds described here show a resemblance in properties to the \HI\
clouds several hundred parsec from the mid plane described by
\citet{lockman2002,lockman2003b,lockman2005}.  The masses, velocities,
and line widths are similar between the samples.  The latitude
distribution of the clouds described here is consistent with a scale
height of the order of $1\ \rm kpc$.  The higher column densities and
smaller diameters for the clouds in the present sample may be a
selection effect because of differences in resolution and sensitivity
between the VLA and the GBT.  The VGPS data may reveal only the
compact, high-column density members of a larger population of
fast-moving clouds, because of the limited sensitivity to low-column
density \HI. On the other hand, the lower resolution of the survey by
\citet{lockman2002} may have diluted the compact high-column density
clouds which are detectable in the VGPS. We speculate that all
fast-moving clouds could be part of a single population which occurs
throughout the disk and up into the halo. An interesting test for this
hypothesis is whether the halo clouds display a similar distribution
in longitude. Also, it is desirable to increase the size of the sample
of detected fast-moving clouds in the disk.  This requires a
combination of high resolution and high sensitivity which can be
achieved with the new ALPHA receiver at the Arecibo radio telescope.

\appendix

\section{Correlation of number of clouds with velocity dispersion}

The line-of-sight velocity of \HI\ at longitude $l$ in the Galactic
plane, following a flat rotation curve with amplitude $V_0$ is
$$
{V_{\rm los}\over V_0 \sin l} = {{1 \over \sqrt{x^2 -2x\cos l +1}}  -1} = F(x)
$$ with $x = d/R_0$ and $d$ is the distance from the Sun. Write
$V_{\rm los}$ as a Taylor series near the tangent point (distance $x_0 =
\cos l$ and $F'(x_0) = 0$):

$$
{V_{\rm los}\over V_0 \sin l} \approx F(x_0) + {1 \over 2} F''(x_0)(x-x_0)^2 + \ldots
$$

Substituting the tangent point velocity 
$$
V_{\rm tan} = V_0 (1 - |\sin l|) {\sin l \over |\sin l|}
$$
which is valid in both the first and fourth Galactic quadrant, we find
$$
V_{\rm los} \approx V_{\rm tan} + {1 \over 2} V_0\ \sin l\ F''(x_0)(x-x_0)^2    \eqno (A1)
$$

The second derivative of $F(x)$ evaluated at the tangent point is
$$
F''(x_0) = - { 1 \over \sin^3 l}
$$ 

\begin{figure}
\centerline{\resizebox{\textwidth}{!}{\includegraphics[angle=0]{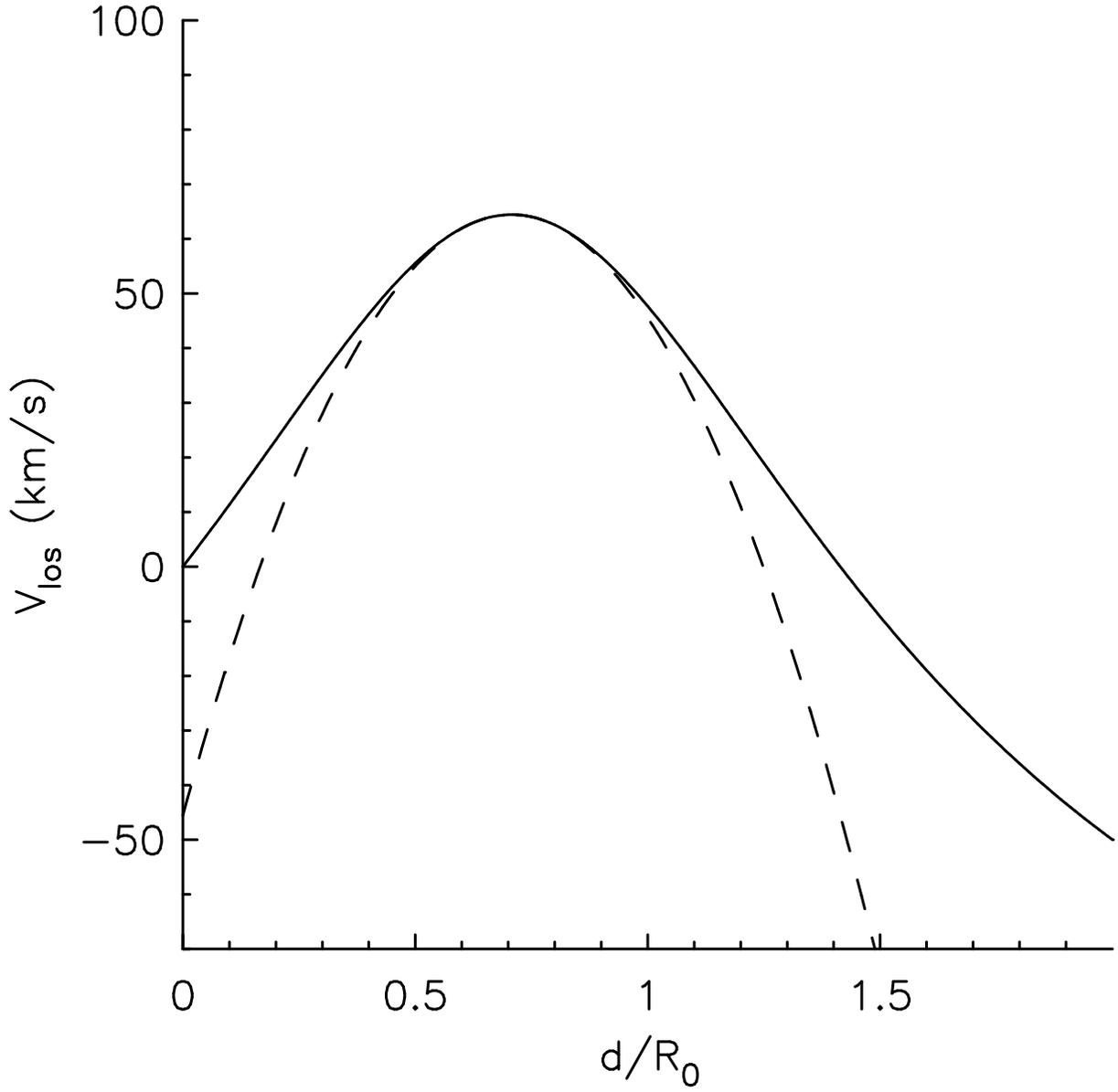}}}
\caption{ Relation between line-of-sight velocity and distance along
the line-of-sight for longitude $l=45^\circ$. (solid curve). The
Taylor expansion from Equation 1 at $l = 45^\circ$ is shown as a
dashed curve. }
\label{Vlos-fig}
\end{figure}

It can be shown that the third derivative $F'''(x) = 0$ if $F'(x) =
0$, so the next term in the Taylor series is of fourth order near the
tangent point.  Figure~\ref{Vlos-fig} shows a graph of $V_{\rm los}$ and the Taylor
expansion following Equation (A1) for longitude $l=45^\circ$.

The number of clouds on the sky $N_c(l)$ is proportional to the number
density of clouds near the tangent point $n_c(x_0)$ and the path
length $\Delta d$ along the line-of-sight over which clouds from a
population with velocity dispersion $\sigma_{\rm c-c}$ are visible at
forbidden velocities.  An implicit integration in the z-direction is
assumed.  We define $\Delta d$ as the line-of-sight distance
corresponding with a velocity $V_{\rm los} = V_{\rm tan} - \sigma_{\rm c-c}$.  In
this case $|x - x_0| R_0 = \Delta d / 2$. Inserting in Equation (A1) yields
$$
\Delta d = \sqrt 8  R_0 |\sin l| \bigg({ \sigma_{\rm c-c} \over V_0} \bigg)^{1 \over 2}
$$
and from this follows for $N_c(l)$
$$
N_c(l) \sim n_c(x_0) |\sin l| \sqrt{\sigma_{\rm c-c}}  \eqno (A2)
$$ 

Note that $n_c(x_0)$ is a function of longitude $l$, but only through
the distribution of clouds with Galactocentric radius. When the radial
distribution of clouds in the Galaxy is specified in the model,
$n_c(x_0)$ is a fixed function of longitude. In practice, clouds with
line-of-sight velocities just beyond the terminal velocity cannot be
counted because of confusion with the wing of the line profile of \HI\
at permitted velocities. This confusion creates an ambiguity in the
determination of $\sigma_{\rm c-c}$, and an underestimation of $N_c(l)$.
The predicted $N_c(l)$ from the models was rescaled to match the
number of clouds in the sample. The effect of this scaling is that
$N_c(l)$ is approximately the same for models with different values of
$\sigma_{\rm c-c}$. At any given longitude, a degeneracy exists between
the normalization of $n_c(x_0)$ and $\sigma_{\rm c-c}$.  Neither of these
quantities can be estimated from the data separately because of
confusion with unrelated emission at the tangent point velocity. The
factor $|\sin l|$ in equation (A2) has the effect that the number of
clouds increases more strongly with $\sigma_{\rm c-c}$ at higher
longitude.  This partially lifts the degeneracy if the longitude
distribution of clouds is considered. However, the sample of clouds
presented in this paper is far too small to constrain the number of
clouds at high longitude to the level required.

Once the longitude distribution is calculated for one value of the
cloud-cloud velocity dispersion $\sigma_{\rm c-c}$, the distribution
for another $\sigma_{\rm c-c}$ is found to a good approximation
through the scaling relationship in Equation (A2).  This scaling is
valid only in the first and fourth quadrants, and for reasonably small
$\sigma_{\rm c-c}$. The models show that this result is valid for the
range of $\sigma_{\rm c-c}$ considered.

\section{Acknowledgements}

The VGPS is supported by a grant to ART from the Natural Sciences and
Engineering Council of Canada.  The authors thank the anonymous
referee for carefully reading the manuscript and for comments which
helped improve the clarity of this paper.

{}

\clearpage
\begin{deluxetable}{lcccccc}
\tabletypesize{\scriptsize}
\tablecolumns{7}
\tablewidth{0pc}
\tablecaption{Observed quantities}
\tablehead{
\colhead{ID} &  \colhead{$V_{\rm LSR}$} &  \colhead{Peak $T_{\rm b}$} & \colhead{$\Delta v$} & \colhead{Peak $N_{\rm HI}$} & 
\colhead{angular diameter} & \colhead{$\int S dv$}\\
\colhead{$l,\  b,\  V_{\rm LSR}$}  & \colhead{($\kms$)}  & \colhead{(K)} & \colhead{($\kms$)} & \colhead{($10^{20}$ cm$^{-2}$)} &
\colhead{(\,$'\ \times\ '$\,)}& \colhead{($\rm Jy\ \kms$)}
} % end tablehead
\startdata                                      
$24.84-0.98+157$      & 157.3 (0.7)    & 10.0 (0.9)     &    16.5 (1.9) & 3.9  & \phn 5  $\times$ 3     & 9.7 (2.9) \\
$25.48+0.16+165$      & 164.6 (0.6)    & 18.7 (1.5)     & \phn8.2 (1.4) & 2.9  & \phn2.9 $\times$ 2.0   & 5.7 (0.7) \\
$27.20-0.82+133$      & 133.0 (0.1)    & 16.8 (1.0)     & \phn5.6 (0.5) & 2.3  & \phn9.1 $\times$ 2.8   & 5.0 (0.3) \\
$28.27+1.05+128$      & 128.1 (0.1)    & 19.2 (1.0)     & \phn5.6 (0.5) & 2.9  & \phn4.8 $\times$ 2.4   & 4.6 (0.3) \\
$28.69-0.09+132$      & 132.3 (0.1)    & 22.5 (1.0)     & \phn5.6 (0.2) & 2.6  & \phn2.0 $\times$ 1.6   & 1.7 (0.3) \\
$28.76+0.58+142$      & 141.6 (0.2)    & 12.2 (1.0)     & \phn5.9 (0.5) & 0.9  & \phn8.7 $\times$ 3.3   & 7.2 (1.1) \\
$29.09+0.18+137$      & 137.0 (0.3)    &  11.0 (1.0)    & \phn7.5 (0.7) & 1.2  & \phn6.5 $\times$ 2.7   & 2.4 (0.4) \\
$30.62-0.57+143$      & 142.8 (0.4)    & 13.9 (2.1)     & \phn5.9 (0.9) & 1.8  & \phn3.7 $\times$ 2.0   & 5.1 (2.4) \\
$32.28-0.72+142$      & 142.1 (0.4)    & \phn8.4 (1.3)  & \phn5.0 (0.9) & 0.8  & \phn4.0 $\times$ 1.8   & 1.3 (0.3) \\
$36.33+0.76+116$      & 115.9 (0.4)    &  \phn7.9 (1.0) & \phn7.5 (1.2) & 1.9  & \phn4.9 $\times$ 3.8   & 8.0 (3.0) \\
$42.42+0.76+101$      & 101.4 (0.2)    &  10.6 (1.0)    & \phn5.0 (0.5) & 1.4  & \phn2.2 $\times$ 1.7   & 1.3 (0.2) \\
$42.46+1.31+105$      & 104.6 (0.5)    & \phn8.6 (1.0)  &    11.7 (1.6) & 1.4  & 17      $\times$ 6     & 27\phn (5)\\
$43.08+0.92+112$      & 111.6 (0.3)    & \phn9.7 (1.2)  & \phn4.7 (0.7) & 1.3  & \phn3.0 $\times$ 1.9   & 2.9 (0.5) \\
$43.36-0.36+97$       & \phn96.6 (0.2) &  13.0 (1.4)    & \phn4.5 (0.5) & 1.5  & \phn4.1 $\times$ 2.9   & 1.0 (0.2) \\
$46.71+1.59+111$      & 111.0 (0.2)    &  12.3 (1.1)    & \phn5.4 (0.5) & 1.4  &  11.8   $\times$ 4.3   & 3.2 (0.6) \\
$59.67-0.39+60$       &  60            &   38           & \phn4.2       & 3.9  & \phn72  $\times$ 12    & 580 (200) \\
$60.70+1.02+58$       &  58            &   23           & \phn3.4       & 2.3  & \phn20  $\times$ \phn6 & \phn49\phn (10)\\
\enddata
\label{observed-tab}
\end{deluxetable}

\begin{deluxetable}{lcccccccc}
\tabletypesize{\scriptsize}
\tablecolumns{9}
\tablewidth{0pc}
\tablecaption{Derived quantities}
\tablehead{
\colhead{ID} & \colhead{d} &  \colhead{z}  & \colhead{$V_{\rm pec}$} & \colhead{diameter} &  
\colhead{$\langle n_{\rm HI}\rangle$} &  \colhead{$M_{\rm HI}$} 
& \colhead{$E_k$ } & \colhead{$T_{\rm kin}$}\\
\colhead{} & \colhead{(kpc)}  & \colhead{(pc)}& \colhead{($\kms$)} &\colhead{(pc $\times$ pc)} &  \colhead{($\rm cm^{-3}$)} & 
 \colhead{($M_\odot$)} &\colhead{($10^{48}$ ergs)} & \colhead{$\rm (10^3\ K)$} } % end tablehead
\startdata
$24.84-0.98+157$    & 7.7  &  $-131$    & 57 &    11 $\times$ \phn6  &     17 &     135  & 4.4  & $<6.0$  \\
$25.48+0.16+165$    & 7.7  &  \phn$+22$ & 55 & \phn6 $\times$ \phn4  &     20 & \phn80   & 2.4  & $<1.5$  \\
$27.20-0.82+133$    & 7.6  &  $-109$    & 29 &    20 $\times$ \phn6  &  \phn7 & \phn68   & 0.6  & $<0.7$  \\
$28.27+1.05+128$    & 7.5  &  $+137$    & 22 &    10 $\times$ \phn5  &     13 &\phn61    & 0.3  & $<0.7$  \\
$28.69-0.09+132$    & 7.5  &  \phn$-11$ & 26 & \phn4 $\times$ \phn3  &     24 &\phn23    & 0.2  & $<0.7$  \\
$28.76+0.58+142$    & 7.5  &  \phn$+76$ & 36 &    19 $\times$ \phn7  &  \phn3 & \phn95   & 0.2  & $<0.8$  \\
$29.09+0.18+137$    & 7.4  &  \phn$+23$ & 34 &    14 $\times$ \phn5  &  \phn5 & \phn31   & 0.4  & $<1.2$  \\
$30.62-0.57+143$    & 7.3  &  \phn$-72$ & 32 & \phn8 $\times$ \phn4  &     11 &  \phn64  & 0.7  & $<0.8$  \\
$32.28-0.72+142$    & 7.2  &  \phn$-90$ & 42 & \phn8 $\times$ \phn3  &  \phn5 & \phn16   & 0.3  & $<0.6$  \\
$36.33+0.76+116$    & 6.8  &  \phn$+90$ & 32 & \phn9 $\times$ \phn7  &  \phn7 & \phn87   & 0.9  & $<1.2$  \\
$42.42+0.76+101$    & 6.3  &  \phn$+84$ & 31 & \phn4 $\times$ \phn3  &     13 &  \phn12  & 0.1  & $<0.6$  \\
$42.46+1.31+105$    & 6.3  &  $+144$    & 35 &    31 $\times$ 11     &  \phn4 &  250     & 3.1  & $<3.0$  \\
$43.08+0.92+112$    & 6.2  &  \phn$+99$ & 47 & \phn5 $\times$ \phn3  &    11  &  \phn26  & 0.6  & $<0.5$  \\
$43.36-0.36+97$     & 6.2  &  \phn$-39$ & 32 & \phn7 $\times$ \phn5  &  \phn8 &\phn\phn9 & 0.1  & $<0.4$  \\
$46.71+1.59+111$    & 5.8  &     $+160$ & 50 &    20 $\times$ \phn7  & \phn4  &\phn25    & 0.6  & $<0.6$  \\
$59.67-0.39+60$     & 4.3  &  \phn$-29$ & 30 &    90 $\times$ 15     & \phn3  &   2500   & 22   & $<0.4$  \\
$60.70+1.02+58$     & 4.2  &  \phn$+75$ & 26 &    24 $\times$ \phn7  & \phn6  &    200   & 1.3  & $<0.3$  \\
\enddata
\label{derived-tab}
\end{deluxetable}

\clearpage
\begin{figure}
%\centerline{\resizebox{\textwidth}{!}{\includegraphics[angle=-90]{f1.eps}}}
\caption{ Distribution of clouds in longitude and velocity.  A $+$
 symbol marks clouds listed in Table~\ref{observed-tab}. The overall
 \HI\ brightness temperature is shown by contours at 5, 10, 20, 40,
 and 80 K, and by a gray scale for $T_{\rm b} \le 10\ \rm K$. The
 smooth curve marks the velocity of the tangent point for a flat
 rotation curve with $V_{\Sun} = 220$ km s$^{-1}$. The dashed line
 marks the boundary of the search area in velocity. The horizontal
 band at $V_{\rm LSR}\approx 120\ \kms$ is the result of low-level
 interference in the zero spacing data.
{\bf [This figure is available as a separate jpg file]}
}
\label{lvmap-fig}
\end{figure}

\begin{figure}
%\centerline{\resizebox{\textwidth}{!}{\includegraphics[angle=0]{f4.eps}}}
\caption{ \HI\ column density maps of clouds listed in Table~1.
Contours are drawn at $-3\sigma$ (dashed), $3\sigma$, $4\sigma$,
\ldots with $\sigma$ the r.m.s. noise in the map ($25.48+0.16+165$:
$\sigma = 3.0 \times 10^{19}\ \rm cm^{-2}$; $28.76+0.58+142$:
$\sigma=1.5\times 10^{19}\ \rm cm^{-2}$; $43.08+0.92+112$:
$\sigma=1.8\times 10^{19}\ \rm cm^{-2}$; $46.71+1.59+111$:
$\sigma=1.8\times 10^{19}\ \rm cm^{-2}$). The gray scales are linear
from $2.5 \times 10^{19}\ \rm cm^{-2}$ to $3.0 \times 10^{20}\ \rm
cm^{-2}$ in all panels. {\bf [This figure is available as a separate jpg file]}
}
\label{NHImap-fig}
\end{figure}

\begin{figure}
%\centerline{\resizebox{\textwidth}{!}{\includegraphics[angle=0]{f7.eps}}}
\caption{ Wide-field \HI\ column density maps of the Galactic plane
with \HI\ emission at forbidden velocities. Panels A and B show two
approximately circular areas where the wing of Galactic \HI\ is
particularly broad. These regions were not included in the search for
clouds. Panel C shows the clouds $59.67-0.39+60$ and
$60.70+1.02+58$. Contrary to the structures in panels A and B, these
clouds do not appear as wings blended with the Galactic profile, but
as separate objects. Gray scales indicate \HI\ column density from $3
\times 10^{19} \ \rm cm^{-2}$ to $3 \times 10^{20} \ \rm cm^{-2}$,
integrated over the velocity range of the feature. Panel A: Velocity
range $128$ to $156\ \kms$. Panel B: Velocity range
$122$ to $149\ \kms$.  Panel C: Velocity range $55$ to $65\ \kms$.
{\bf [This figure is available as a separate jpg file]}
}
\label{group-fig}
\end{figure}

\begin{thebibliography}{}

\bibitem[Bregman(1980)]{bregman1980} Bregman, J. N. 1980, \apj, 236, 577

\bibitem[Burton \& Gordon(1978)]{burtongordon78} Burton, W. B., \& 
Gordon, M. A. 1978, \aap, 63, 7

\bibitem[Clemens(1985)]{clemens} Clemens, D.P. 1985, \apj, 295, 422

\bibitem[Dame(1993)]{dame1993} Dame, T. M. 1993, in 'Back to the Galaxy', ed. S.S. Holt and F. Verter, AIP Press: New York, p. 267

\bibitem[De Heij et~al.(2002)]{deheij2002} 
De Heij, V., Braun, R., \& Burton, W. B. 2002, \aap, 392, 417

\bibitem[Dickey \& Lockman(1990)]{dickeylockman} Dickey, J. M., \& Lockman, F. J. 1990, \araa, 28, 215

\bibitem[Dwarakanath(2004)]{dwark} Dwarakanath, K.S. 2004, 
Bull. Astr. Soc. India, 32, 215 

\bibitem[Dwarakanath et~al.(2004)]{dwarakanath2004b} Dwarakanath, K.S., Goss, W. M., Zhao, J. H., \& Lang, C. C. 2004, J. Astr. Ap., 25, 129

\bibitem[Englmaier \& Ortwin(1999)]{englmaier1999} Englmaier, P., \& Ortwin, G. 1999, \mnras, 304, 512

\bibitem[Green(2004)]{green2004} Green, D. A. 2004, A Catalogue of Galactic Supernova Remnants (2004 January version), Mullard Radio Astronomy Observatory, Cavendish Laboratory, Cambridge, UK (available on the World-Wide-Web at ``{\tt http://www.mrao.cam.ac.uk/surveys/snrs/}'')

\bibitem[Hartmann \& Burton(1997)]{hartmann} Hartmann D., Burton W.B. 1997, Atlas of galactic neutral hydrogen, Cambridge University Press

\bibitem[Higgs et~al.(2001)]{higgs2001} Higgs, L. A., Kerton, C. R., \& Knee, L. B. G. 2001, \aj, 122, 3155 

\bibitem[Kalberla et~al.(1998)]{kalberla98} Kalberla, P.M.W., 
Westphalen, G., Mebold, U., Hartmann, D., \& Burton, W.B. 1998, \aap, 332, L61

\bibitem[Kerton et~al.(2002)]{kerton2002} Kerton, C. R., Knee, L. B. G., Higgs, L. 2002, in ASP Conf. Ser. 276, Seeing Through the Dust: The Detection of HI and the Exploration of the ISM in Galaxies, ed. A. R. Taylor, T. L. Landecker \& A. G. Willis, (San Francisco: ASP), 136


\bibitem[Koo et~al.(1990)]{koo1990} Koo, B.-C., Reach, W. T., Heiles, C., Fesen, R. A., \& Shull, J. M. 1990, \apj, 364, 178

\bibitem[Koo \& Heiles(1991)]{koo1991} Koo, B.-C., \& Heiles, C., 1991, \apj, 382, 204

\bibitem[Koo \& Kang(2004)]{koo2004} Koo, B.-C., Kang, J.-H. 2004, \mnras, 349, 983

\bibitem[Kulkarni \& Fich(1985)]{kulkarni1985} Kulkarni, S. R., \& Fich, M. 1985, \apj, 289, 792

\bibitem[Lockman(1989)]{lockman1989} Lockman, F. J. 1989, \apjs, 71, 469 

\bibitem[Lockman et~al.(1996)]{lockman1996} Lockman, F. J., Pisano, D. J., \& Howard, G. J. 1996, \apj, 472, 173 

\bibitem[Lockman(2002)]{lockman2002} Lockman, F. J. 2002, \apj,  580, L47

\bibitem[Lockman \& Stil(2003)]{lockman2003a} Lockman, F. J., \& Stil, J. M. 2004, in ASP Conf. Ser. 317, Milky Way Surveys: The Structure and Evolution of Our Galaxy, ed. D. Clemens, R. Y. Shah, \& T. Brainerd, (San Francisco: ASP), 20 (astro-ph/0311047)

\bibitem[Lockman(2003)]{lockman2003b} Lockman, F. J. 2003, IAU symp Vol. 217, 2004, astro-ph/0311217

\bibitem[Lockman \& Pidopryhora (2005)]{lockman2005} Lockman, F. J., \& Pidopryhora, Y.  2005 in in ASP Conf. Ser. 331, Extra-Planar Gas, ed. R. Braun, (San Francisco: ASP), 59

\bibitem[Mohan et~al.(2004)]{mohan} Mohan, R., Dwarakanath, K.S., \& Srinivasan, G. 2004, J. Astr. Ap., 25, 185 

\bibitem[M\"{u}nch \& Zirin (1961)]{munchzirin} M\"{u}nch, G., \& Zirin, H. 1961, \apj, 133, 11

\bibitem[Radhakrishnan \& Srinivasan(1980)]{radhakrishnan1980}
Radhakrishnan, V., \& Srinivasan, G. 1980, J. Astr. Ap., 1, 47

\bibitem[Siluk \& Silk(1974)]{siluk1974} Siluk, R. S., \& Silk, J. 1974, \apj, 192, 51

\bibitem[Taylor et~al.(2002)]{taylor2002} Taylor, A. R., Stil, J. M., Dickey, J. M., McClure-Griffiths, N. M., Martin, P. G., Rothwell, T., Lockman, F. J. 2002, in ASP Conf. Ser. 276, Seeing Through The Dust: The Detection Of \HI And The Exploration Of The ISM In Galaxies, ed. A. R. Taylor, T. L. Landecker, \& A. G. Wills (San Francisco: ASP), 68

\bibitem[Weiner \& Sellwood(1999)]{weiner} Weiner, B.J., \& Sellwood, J.A. 2003, \apj, 524, 112

\bibitem[Welsh et~al.(2004)]{welsh2004} Welsh, B. Y., Sallmen, S., \& Lallement, R. 2004, \aap, 414, 261

\end{thebibliography}
\end{document}